\documentclass[
preprint, 
 amsmath,amssymb,
 aps, 
]{revtex4-2}

\usepackage{graphicx}
\usepackage{dcolumn}
\usepackage{bm}
\newcommand{\bg}{\begin{eqnarray}}
\newcommand{\ed}{\end{eqnarray}}
\newcommand{\oprzerotwo}{
(
\gamma^0\partial_0 - \gamma^2\partial_2
)
}
\newcommand{\barpar}
{
\partial\hspace{-0.22 cm}\slash
}
\newcommand{\mass}{
\frac{mc}{\hbar}
}
\newcommand{\hatbp}
{
\hat{\partial}\hspace{-0.23 cm}\slash
}
\newcommand{\dira}{
(i\barpar -m)\varphi = 0
}
\newcommand{\dirm}{
(i\barpar -m)(\apfi +\Delta \apfi) = 0,~\text{neglect of}~O(m^4)
}
\newcommand{\kleigo}{
(\square^2 + m^2)\varphi = 0
}
\newcommand{\apfi}
{
\tilde{\varphi}
}
\newcommand{\rweight}
{
m^3\left(x_0 + x_2 + c_0\right)^{m}
}
\newcommand{\weight}
{
m^4\left(x_0 + x_2 + c_0\right)^{m}
}
\newcommand{\dweight}
{
\frac{m^5}
{
\left(x_0 + x_2 + c_0\right)^{1-m}
}
}
\newcommand{\hattr}{
\hat{\partial}_{\text{tr}}\hspace{-0.45 cm}\slash
}

\newcommand{\hhoed}{
\hat{\hat{\barpar}}
}
\newcommand{\dhoed}{
\hat{\hat{\barpar}}^2
}
\newcommand{\oper}{
 (\partial_1 - \gamma^1\gamma^3\partial_3)
}

\newcommand{\seeup}{
C'_\upsilon
}
\newcommand{\sedown}{
C'_\delta
}
\newcommand{\seupo}
{
C'_{\upsilon,1}
}
\newcommand{\seupt}
{
C'_{\upsilon,2}
}
\newcommand{\sedoo}
{
C'_{\delta,1}
}
\newcommand{\sedot}
{
C'_{\delta,2}
}
\newcommand{\hhga}{
\hat{\hat{\gamma}}
}

\begin{document}


\title{Are neutrinos changed in the cross-hairs of two metric tensors?}

\author{Han Geurdes}
 \altaffiliation[Also at ]{GDS applied mathematics BV Netherlands.}
 \email{Contact author: han.geurdes@proton.me}
\affiliation{%
 \\
  Pensioner C vd Lijnstraat 164 2593 NN, Netherlands
}

\date{\today}

\begin{abstract}
A restriction was found in the mathematics of the Dirac equation for a free neutrino type of particle.
The basic assumption here is the equivalence of the four variables of spacetime.
A perspective is defined as a metric tensor format.
We asked what happens when we add a perspective where a time variable, $\sim ct$ unit [meter], becomes a space variable, unit [meter], and vice versa. 
A Lorentz invariant mixed metric tensor equation can be set up in an attempt to describe the neutrino in the cross-hairs of two different perspectives.
Despite the equivalence of variables, and despite the fact that no single perspective is likely to be preferred, we find that 
the neutrino is internally changed in the cross-hairs of two different perspectives.
Of course, if the two perspectives do not go together this conclusion collapses.
\end{abstract}

\maketitle

\section{\label{intro}Introduction}
The Dirac equation was a fruitful approach to the discovery of new states of matter \cite{Ander}.
A key assumption in the present work is that the four variables in special relativistic spacetime are equivalent. 
This could mean that one can select a variable and denote it as time related and the other three variables as space.
Two different perspectives may have different selections of variables.
Spacetime is not supposed to be affected by how we select our variables in it to describe it. 
It is also not obliged to select our perspective or only have one. 
If that's the case, then in the words of the philosopher P. Feyerabend \cite{Fey}, "anything" goes.
Hence, there is the possibility that, then, a neutrino ends up in the cross-hairs of both perspectives.
This induces a restriction in the form of a reduced zero mass Dirac equation.
\subsection{Preliminaries}
In the paper the mathematical structure of the Dirac equation approach to relativistic quantum mechanics is studied.
Let us start by introducing three forms of differential operators.
The first one is the usual operator. In Feynman's dagger format,
$\barpar=\gamma^0\partial_0 - \gamma^1\partial_1-\gamma^2\partial_2-\gamma^3\partial_3$.
The $\gamma$ matrices are
\bg\label{gg}
\gamma^0=\left(\begin{array}{cc}1 & 0 \\ 0 & -1\end{array}\right), \,
\gamma^k=\left(\begin{array}{cc}0 & \sigma_k \\ -\sigma_k & 0\end{array}\right)
\ed
with $k=1,2,3$. 
Here, the matrices are: $1$ is $2\times 2$ unity matrix, $0$ is the $2\times 2$ zero matrix and the others are
\bg\label{gg1}
\sigma_1=\left(\begin{array}{cc}0 & 1 \\ 1 & 0\end{array}\right), 
\sigma_2=\left(\begin{array}{cc}0 & i \\ -i & 0\end{array}\right), \text{and}, 
\sigma_3=\left(\begin{array}{cc}1 & 0 \\ 0 & -1\end{array}\right)
\ed
Such a definition of Pauli matrices is perfectly allowed.
The $\barpar$ corresponds with the metric tensor signature $(+,-,-,-)$.
Another approach is based on a different set of Dirac-Clifford matrices denoted by $\hat{\gamma}$.
We have, $\hatbp=\hat{\gamma}^2\partial_2 - \hat{\gamma}^0\partial_0-\hat{\gamma}^1\partial_1-\hat{\gamma}^3\partial_3$.
Here, $\hat{\gamma}^2 = i\gamma^2$ and $\hat{\gamma}^0 = i\gamma^0$, while, $\hat{\gamma}^1=\gamma^1$ and $\hat{\gamma}^3=\gamma^3$.
The third one has two spatial variables exchanging position with respect to $\barpar$. It is $\hat{\hat{\barpar}}=\gamma^0\partial_0  -\gamma^1\partial_2-\gamma^2\partial_1 -\gamma^3\partial_3$.

Let us consider this last one first.
Now it is clear that $(i\hhoed +m)(\hhoed -m)\xi(x)=(-\square^2 - m^2)\xi(x)$.
If $(-\square^2 - m^2)\xi(x) = 0$, $\xi$ a $4\times 1$ vector, because $(i\barpar -m)\xi(x)=0$, then $\hhoed$ and $\barpar$ are not essentially different on a Klein-Gordon level.
Here we note $\square^2 = \partial_0^2 -\nabla^2$.
If, however, $\hatbp ^2$ is inspected we see $\hatbp ^2=\partial_2^2-\partial_0^2-\partial_1^2-\partial_3^3$. This is different from ${\dhoed}$ and ${\barpar}^2$, with ${\dhoed}$ = ${\barpar}^2$.

The $\hatbp$ operator is related to a metric. 
From the original point of view and signature $(+,-,-,-)$, it has a \emph{new} signature $(-,-,+,-)$.
The Dirac-Clifford matrices with a hat are, $\hat{\gamma}^\mu = i\gamma^\mu$ for $\mu=0,2$ and $\hat{\gamma}^\mu =\gamma^\mu$ for $\mu=1,3$. 
It follows an idea employed in spin theory in theoretical chemistry \cite{Bryan}.
The closer look into the role of the metric tensor is also inspired by the work of Sakharov \cite{Sach}.

One can look at it as if the $(+,-,-,-)$ variable $x_2$ becomes a time-variable in $(-,-,+,-)$; i.e. a space variable becomes a time-form variable in the system with a hat. 
And the $(+,-,-,-)$ variable $x_0$ becomes a space-form variable in $(-,-,+,-)$. 
Spacetime isn't changed by changing the role of the variables from $(+,-,-,-)$ to $(-,-,+,-)$.
Therefore, the two approaches are equivalent considering their description of "events" in spacetime.
Moreover, in the case of $\hat{\gamma}$,  the basic relations between the $\gamma$ matrices remain, $\{\gamma^\mu,\gamma^\nu\}=2g^{\mu,\nu}$. 
And so, $\{\hat{\gamma}^\mu,\hat{\gamma}^\nu\}=2\hat{g}^{\mu,\nu}$ with $\hat{g}$ with signature $(-,-,+,-)$ when we look at it from a $g^{\mu,\nu}$ perspective. 
It is clear that $(\hat{\gamma}^2)^2=1$, i.e. the matrix pre-multiplication for a  time variable. 
And $(\hat{\gamma}^0)^2=-1$ the matrix pre-multiplication for a space variable.
\subsection{\label{sec:level2}The perspective with the hat and Lorentz}
A perspective is defined as a particular metric tensor.
Let us start by looking at a free particle Dirac equation in $g^{\mu,\nu}$ signature $(+,-,-,-)$, with,  $\dira$.
We will consider a small mass, and, free fermion particle like e.g. a neutrino.
In the $\hat{g}^{\mu,\nu}$ metric, with signature, $(-,-,+,-)$ it is possible to define the operator, $i\hatbp +m$.  
This gives the possibility to write a kind of mixed form Klein-Gordon equation,
\bg\label{eqbas}
(i\hatbp + m)(i\barpar -m) \varphi (x) = 0
\ed
This equation is inspected for its behavior under Lorentz transformation.
The $m$ in (\ref{eqbas}) is in fact $\mass$.
For an ultra light neutrino we can have, because neutrino mass is supposed to be less than around the order of $10^{-37}$ [kg], for instance $m\sim 10^{-46} \times 10^8\times  10^{34} =10^{-4}$.

Let's follow the basics of Lorentz transformation in Bjorken and Drell \cite{Bjork}.
In the first place we have, starting with $x=(x_0,x_1,x_2,x_3)$, in the new coordinates the vector $x'=ax$, with, $x'_\nu = a_\nu^\mu x_\mu$.
The $4\times 4$ matrix $a$ represents the Lorentz transformation of coordinates $x$.
Secondly, there is a transformation matrix ($4\times 4$) such that, $S(a)\varphi(x)=\varphi^{\prime}(x')$ and we also see $S(a)S^{-1}(a)=S^{-1}(a)S(a)=1$.
Therefore, equation (\ref{eqbas}) can, following \cite{Bjork}, be transformed to 
\bg\label{eqlor}
(iS(a)\hat{\gamma}^\sigma a_\sigma^\lambda  S^{-1}(a) \partial^{\,\prime}_\lambda + m)(iS(a)\gamma^\mu a^\nu_\mu  S^{-1}(a)\partial^{\,\prime}_\nu -m) \varphi^{\prime} (x') = 0
\ed
And because $\gamma^\nu = S(a)\gamma^\mu a^\nu_\mu  S^{-1}(a)$, it is found, in conjunction with the definition of the $\hat{\gamma}$ matrices, that $(i\hatbp^{\,\prime} + m)(i\barpar^{\,\prime} -m) \varphi^{\prime} (x') = 0$. 
The description with both $g^{\mu,\nu}$ and   
$\hat{g}^{\mu,\nu}$  in the mixed equation (\ref{eqbas}) is a Lorentz invariant expression.
\subsection{The perspective with the two hats and Lorentz}
Let us also look at the Lorentz transformation of (\ref{eqbas}) where $\hatbp$ is replaced by $\hhoed$.
It is noted that both $\hhoed $ and $\barpar$ lead to the same Klein-Gordon equation when $\dira$.
Therefore if we have $x'=ax$ and $a$ a Lorentz transformation matrix, in both cases we must have $S^{-1}(a)\varphi'(x')=\varphi(x)$.
Then $(i\hhoed + m)(iS(a)\gamma^\mu a_\mu^\nu S^{-1}(a)\partial'_\nu - m)\varphi'(x') = 0$ follows.
This implies
\bg\label{dhat1}
(iS(a){\hhga}{\,}^\lambda a_\lambda ^\kappa S^{-1}(a)\partial'_\kappa + m)(iS(a)\gamma^\mu a_\mu^\nu S^{-1}(a)\partial'_\nu - m)\varphi'(x') = 0
\ed
Here, $\hhga{\,}^2= \gamma^1$ and $\hhga{\,}^1 = \gamma^2$ and $\hhga{\,}^3= \gamma^3$ and $\hhga{\,}^0= \gamma^0$.
For a Lorentz transformation of $\dira$ we know \cite{Bjork} that $S(a)\gamma^\mu a_\mu^\nu S^{-1}(a) = \gamma^\nu$.
And it then implies that $S(a){\hhga}{\,}^\lambda a_\lambda ^\kappa S^{-1}(a)={\hhga}{\,}^\kappa$ if (\ref{dhat1}) is to be Lorentz invariant as well.
This necessitates a.o.
\bg\label{dhat2}
\begin{array}{cc}
S(a)\left[\gamma^0a_0^1-\gamma^1 a_1^1-\gamma^2 a_2^1-\gamma^3 a_3^1\right]S^{-1}(a)=\gamma^1 & \text{normal case}
\\
S(a)\left[\gamma^0a_0^2-\gamma^1 a_2^2-\gamma^2 a_1^2-\gamma^3 a_3^2\right]S^{-1}(a)=\gamma^1 & \text{double hat}
\\
S(a)\left[\gamma^0a_0^2-\gamma^1 a_1^2-\gamma^2 a_2^2-\gamma^3 a_3^2\right]S^{-1}(a)=\gamma^2 & \text{normal case}
\end{array}
\ed
And so, subtracting the first two equations of (\ref{dhat2}) and making use of the third one
\bg\nonumber
S(a)\left[\gamma^0 (a_0^2 -a_0^1)-\gamma^1(a_1^2+a_2^2 - a_1^1-a_1^2)-\gamma^2(a_2^2+a_1^2- a_2^1-a_2^2)-\gamma^3 (a_3^2-a_3^1)\right]S^{-1}(a)=0
\ed
This gives
\bg\label{dhat3}
\gamma^2 - \gamma^1 -(a_2^2 - a_1^2 )S(a)\left(\gamma^1 - \gamma^2\right)S^{-1}(a)=0 
\ed
The equation (\ref{dhat3}) cannot be generally true because when $a_2^2 - a_1^2 = 0$, which is a genuine possibility, we need to have, $\gamma^2 = \gamma^1$.
The latter is not true.
The $\hhoed$ and $\barpar$ equation cannot mix like with $\hatbp$ and $\barpar$ in an equation such as (\ref{eqbas}).
\section{\label{sec:level2.1}Klein - Gordon like in equation  (\ref{eqbas})}
Let us, in a particular $(+,-,-,-)$ coordinate frame, process the equation (\ref{eqbas}), i.e. the equation $(i\hatbp + m)(i\barpar -m) \varphi (x) = 0$.
This is possible because $\hat{\gamma}$ matrices are a linear transformation of $\gamma$ matrices.
We then have, 
\bg\label{KG1}
[-\hatbp\barpar - m^2 -im(\hatbp - \barpar)]\varphi = 0
\ed
Because of $\{\gamma^\mu, \gamma^\nu \}=2g^{\mu,\nu}$
it is subsequently found that
$-\hatbp\barpar = i(\partial_0^2-\partial_2^2)+(\partial_1^2+\partial_3^2)$.
To continue, because of $\dira$, and $\{\gamma^\mu, \gamma^\nu \}=2g^{\mu,\nu}$, we also have $\kleigo$. 
Because $\square^2 = \partial_0^2 - \nabla^2$ and $\nabla^2 = \partial_2^2 +(\partial_1^2 + \partial_3^2)$, it follows that 
$[(\partial_1^2+\partial_3^2) -m^2]\varphi = (\partial_0^2 -\partial_2^2)\varphi$.
This in turn implies that $[-\hatbp\barpar - m^2]\varphi = (i+1)(\partial_0^2-\partial_2^2)\varphi$.

Now looking at $im(\hatbp - \barpar)$, the operator part is $\hatbp - \barpar = (i+1)(\gamma^2\partial_2 - \gamma^0\partial_0)$.
Note that, $\partial_0^2-\partial_2^2=(\gamma^2\partial_2-\gamma^0\partial_0)^2$.
This implies that equation (\ref{KG1}) can be rewritten like
\bg\label{KG2}
(\gamma^2\partial_2-\gamma^0\partial_0)[(i+1)(\gamma^2\partial_2-\gamma^0\partial_0)+(1-i)m ]\varphi = 0
\ed
The $(-i)(1+i)=(1-i)$.
Obviously we could also have written equation (\ref{KG2}) in the form, $[(i+1)(\gamma^2\partial_2-\gamma^0\partial_0)+(1-i)m ](\gamma^2\partial_2-\gamma^0\partial_0)\varphi = 0$.
Let's continue, however, with (\ref{KG2}).
Because of $\dira$, we have $[i(\gamma^2\partial_2-\gamma^0\partial_0) + m]\varphi = -i(\gamma^1\partial_1 + \gamma^3\partial_3)\varphi$. 
And via $[(\gamma^2\partial_2-\gamma^0\partial_0) -i m]\varphi = -(\gamma^1\partial_1 + \gamma^3\partial_3)\varphi$, it follows from (\ref{KG2}),
\bg\label{KG3}
(\gamma^2\partial_2-\gamma^0\partial_0)(\gamma^1\partial_1 + \gamma^3\partial_3)\varphi = 0
\ed
And so, $\gamma^1 (\gamma^0\partial_0-\gamma^2\partial_2)(\partial_1 -\gamma^1 \gamma^3\partial_3)\varphi = 0$.
If we write $\phi_{1,3}=(\partial_1 -\gamma^1 \gamma^3\partial_3)\varphi$ then
\bg\label{KG4}
\gamma^0\partial_0 \phi_{1,3} = \gamma^2\partial_2 \phi_{1,3}
\ed
Equation (\ref{KG3}) is a consequence of the cross-hairs (\ref{eqbas}).
\subsection{Charge conjugation $\mathcal{C}$}\label{cconj}
Before going into the details of an approximate solution, let us inspect the effect of the charge conjugation matrix. 
We define the charge conjugation matrix, $\mathcal{C} = i\gamma^2\gamma^0$ we have $\mathcal{C}^{-1}\gamma^\tau \mathcal{C}=-\gamma^\tau$ when $\tau=0,2$, and $\mathcal{C}^{-1}\gamma^\tau \mathcal{C}=\gamma^\tau$, when $\tau=1,3$, like in \cite{Bjork}. 
Then when we define $\psi(\xi) = \mathcal{C}^{-1}\varphi(x_0,x_1,x_2,x_3)$ and $\xi = (ix_0,x_1,ix_2,x_3)$ it follows that 
\bg\label{psihat}
(i\hat{\gamma}^2\partial_{\xi_2}-i\hat{\gamma}^0\partial_{\xi_0}-i{\gamma}^1\partial_{\xi_1}-i{\gamma}^3\partial_{\xi_3} - m)\psi(\xi) = 0
\ed
The equation (\ref{psihat}) is not employed explicitly in the paper but it is noted that in some respect, charge conjugation gives a similar sort of Diracr equation.
However, do note that $\xi_0$ and $\xi_2$ are imaginary variables whereas in our analysis we have $x_0$ and $x_2$ both real. When a reduced zero mass Dirac equation exists in $x_0$ and $x_2$ only, premultiplication with $\mathcal{C}$ and interchange  of variables will give an equation in $\hat{g}$. 
\section{A neglect of $O(m^4)$ solution to $\dira$}\label{Odri3}
This allows a neglect of term of the order of $O(m^4)\sim 10^{-16}$ approximate solution of the free Dirac equation $\dira$ to be associated to physics.
Remember $m$ represents $\mass$.
Let us then define the approximation $4\times 1$ vector $\apfi$ as
\bg\label{ap1}
\apfi = 
\left[
1-\frac{im}{4}\left(
\gamma^0 x_0 + \gamma^1 x_1 + \gamma^2 x_2 + \gamma^3 x_3 
\right)
-\frac{m^2}{8}x^2
+\frac{im^3}{24}
\left(
\gamma^0 x_0^3
-
\sum_k \gamma^k x_k^3
\right)
\right]
C
\ed
with $C$ a $4\times 1$ vector, constant in $x$. 
Furthermore, spacetime coordinates are represented by  $x=(x_0,x_1,x_2,x_3)$  and $x^2 = x_0^2 - (x_1^2+x_2^2+x_3^2)$ in $(+,-,-,-)$.
From (\ref{ap1}) it follows that $\partial_0^2 \apfi = \left(-\frac{m^2}{4}+\gamma^0\frac{im^3}{4}x_0\right)C$. 
Furthermore, $\partial_1^2 \apfi =\left(\frac{m^2}{4}-\gamma^1\frac{im^3}{4}x_1\right)C$, so that, 
\bg\label{klgo1}
(\partial_0^2-\nabla^2)\apfi = 
-m^2\left[
1-\frac{im}{4}
(\gamma^0x_0+\sum_k \gamma^k x_k)
\right]C
\ed
This is at least a neglect of $O(m^2)$ terms in $\apfi$ approximation with $(\square^2 + m^2)\apfi =0$.
Hence, it is $(\partial_0^2-\nabla^2)\apfi = 
-m^2\left[
1-\frac{im}{4}
(\gamma^0x_0+\sum_k \gamma^k x_k)
\right]C$
And so, 
$
(\partial_0^2-\nabla^2)\apfi = 
-m^2\apfi
$ with neglect of $O(m^2)$ in $\apfi$.

Further, the $\apfi$ from (\ref{ap1}) is a neglect of $O(m^4)$ terms approximation of $\varphi$ in the Dirac equation $\dira$, because, in $O(m^4)$,
\bg\label{ap2}
\begin{array}{c}
i\gamma^0\partial_0\apfi = 
(i\gamma^0)
\left[
-\frac{im}{4}\gamma^0 - \frac{m^2}{4}x_0
+\gamma^0\frac{im^3}{8}x_0^2
\right]
C
=m
\left[
\frac{1}{4} -\frac{im}{4}\gamma^0 x_0
-
\frac{m^2}{8}x_0^2\right]
C
\\
\\
-i\gamma^1\partial_1\apfi = (-i\gamma^1)
\left[
-\frac{im}{4}\gamma^1 + \frac{m^2}{4}x_1
-\gamma^1
\frac{im^3}{8}x_1^2
\right]
C
=m
\left[
\frac{1}{4} -\frac{im}{4}\gamma^1 x_1
+
\frac{m^2}{8}x_1^2
\right]
C
\\
\\
-i\gamma^2\partial_2\apfi = (-i\gamma^2)
\left[
-\frac{im}{4}\gamma^2 + \frac{m^2}{4}x_2
-\gamma^2
\frac{im^3}{8}x_2^2
\right]
C
=m
\left[
\frac{1}{4} -\frac{im}{4}\gamma^2 x_2
+
\frac{m^2}{8}x_2^2
\right]
C
\\
\\
-i\gamma^3\partial_3\apfi = (-i\gamma^3)
\left[
-\frac{im}{4}\gamma^3 + \frac{m^2}{4}x_3
-\gamma^3
\frac{im^3}{8}x_3^2
\right]
C
=m
\left[
\frac{1}{4} -\frac{im}{4}\gamma^3 x_3
+
\frac{m^2}{8}x_3^2
\right]
C
\end{array}
\ed
Therefore, we can conclude that in neglecting $O(m^4)$ the $\apfi$ solves $\dira$.
It is exact to the neglect of the order of terms with $10^{-16}$ and normalizable for $x_\mu \in  [-L_\mu/2,L_\mu/2] $, with $L_\mu/2$ relatively small real positive.
\subsection{Disturbance $\Delta \apfi$}
Subsequently, let us introduce a $\Delta \apfi$ in the definition of $\apfi$ such that the previous analysis remains valid and we have a nontrivial equation (\ref{KG4}).
The first step is to have $\Delta \apfi \propto \weight$.
The constant $c_0$ independent of $x$ is $> (L_0+L_2)/2$. 
This is so because a $\partial_0$ or $\partial_2$ on $\Delta \apfi$ would then be of the order of neglect of $O(m^4)$ for $x$ in a proper domain.
The second step is to make use of pseudo-scalar $2\times 2$ multiplication. 
In particular e.g. multiplication with $1\pm \sigma_1$ times a $4\times 4$ matrix is via $4\times 4$ unity matrix
\bg\label{ps1}
\left(
\begin{array}{cc}
1\pm \sigma_1 & 0\\
0 & 1\pm \sigma_1
\end{array}
\right)
\ed
The 1 here is the $2\times 2$ unity matrix. The $0$ is the $2\times 2 $ zero matrix.

Subsequently, a $\Delta \apfi$ can be defined as
\bg\label{aps2}
\Delta \apfi=
\rweight
\left[
(1-\sigma_1)x_1 - (1+\sigma_1)x_3
\right]
\Gamma'C'
\ed
Here, the $\Delta\apfi = O(m^3)$.
Therefore it fits the neglect of $O(m^4)$ approximation of $\dira$.
The $x_0$ or the $x_2$ derivative of $m\times\rweight$ is $\dweight$. 
Because $m\sim 10^{-4}$, and 
$x_\nu \in \mathcal{D}=\bigcup_{\mu=0}^3 \{x_\mu\in\mathbb{R}\,|\, (-L_\mu/2)\leq x_\mu\leq L_\mu /2\}$ 
the $(x_0+x_2+c_0)^m$ part is close to unity.
The $\apfi + \Delta\apfi$ is similarly normalized as $\apfi$ when all $L_\mu$ are finite relatively small.
In addition the $C'$ is a $4\times 1$ vector constant in $x$.
The $4\times 4$ matrix $\Gamma'$ is defined by
\bg\label{ps3}
\Gamma'=
\left(
\begin{array}{cc}
-\mathcal{G} & 0
\\
0 & \mathcal{G} 
\end{array}
\right)
\ed
and $\mathcal{G}= \left(\frac{\sigma_1 - \sigma_3}{2}\right)(1+\sigma_1)$.
\subsection{A closer look at $\dirm$}\label{oprzeto}
We find that
\bg\label{df2}
\left(-i\gamma^1\partial_1 -i\gamma^3\partial_3\right)\Delta \apfi
=
(-i)\rweight
\left(
\gamma^1
(1-\sigma_1)
-
\gamma^3
(1+\sigma_1)
\right)
\Gamma'C'
\ed
Now, $\gamma^1
(1-\sigma_1) = (1-\sigma_1) \gamma^1$
and
$\gamma^3
(1+\sigma_1) =
(1-\sigma_1)\gamma^3$.
With the definition of $\Gamma'$ in (\ref{ps3}), the subsequent relevant evaluation of (\ref{df2}) is
\[(1-\sigma_1)(\sigma_1-\sigma_3)\left(\frac{\sigma_1-\sigma_3}{2}\right)(1+\sigma_1) = (1-\sigma_1)(1+\sigma_1) = 0.\]
The $\sigma_1\sigma_3 + \sigma_3\sigma_1 = 0$.
And so we can conclude that 
$\left(-i\gamma^1\partial_1 -i\gamma^3\partial_3\right)\Delta \apfi = 0$ such as required to obtain the solution of the free Dirac equation $\dirm $, in the previous section.
The analysis is maintained because the $m^4$ in the $\partial_0$ and $\partial_2$ will drop off at a certain stage considering (\ref{KG4}).
\subsection{The $x_0$ and $x_2$ dependence}
Because of separable $x$ variables in $\apfi$, in (\ref{ap1}), we have
$\oprzerotwo  \oper\apfi = 0$.
It is because, $\oper \apfi$ is independent of $x_0$ and $x_2$.
Furthermore $m\Delta \apfi$ is $O(m^4)$. 
And so, $m\sigma_1\Delta\apfi$ is $O(m^4)$. 
Hence, we may consider
\bg\label{eq021}
\oprzerotwo  \oper(\apfi+m\sigma_1\Delta\apfi) = 0
\ed
And so, it follows that
$\oprzerotwo  \oper(\sigma_1\Delta\apfi) = 0$.
Now, we have $\sigma_1(1-\sigma_1)=-(1-\sigma_1)$, while, $\sigma_1(1+\sigma_1)=(1+\sigma_1)$.
This, with $\oper = (-\gamma^1)(\gamma^1\partial_1+\gamma^3\partial_3)$ leads to 
\bg\label{tdep4}
\sigma_1(1-\sigma_1)(\sigma_1 + \sigma_3) \left( \frac{\sigma_1 - \sigma_3}{2}\right)\sigma_1(1+\sigma_1)=
\sigma_1(1-\sigma_1)\sigma_3(1+\sigma_1)=
\\\nonumber
=
2\sigma_1\sigma_3 (1+\sigma_1) \not \equiv 0
\ed
This, in turn, implies that 
\bg\label{tdep4a}
\oper(\sigma_1\Delta \apfi) = 2\rweight\Sigma_1 C'
\ed
in principle can be non-vanishing. 
Here,
\bg\label{tdep5}
\Sigma_1=
\left(
\begin{array}{cc}
0 & \sigma_1\sigma_3 (1+\sigma_1)
\\
\sigma_1\sigma_3 (1+\sigma_1) & 0
\end{array}
\right)
\ed
The $\Sigma_1$ structure arises from $\left(1-\sigma_1\right)
\left(
\gamma^1+\gamma^3
\right)
\Gamma'$, the result in (\ref{tdep4}) and the definition in (\ref{ps3}). 
Because $\sigma_1(1+\sigma_1)=(1+\sigma_1)$, it is found that, $\sigma_1\sigma_3 (1+\sigma_1) = -\sigma_3 (1+\sigma_1)$.
Note also that $\oper(\sigma_1\Delta \apfi)$ depends on $x_0$ and $x_2$.
\subsection{Continuation with equation (\ref{eq021}) }
Let´s look at equuation $\oprzerotwo  \oper(\sigma_1\Delta\apfi) = 0$ with the knowledge we gained in the previous subsection-\ref{oprzeto}.
\bg\label{fftjes1}
\begin{array}{c}
\gamma^0\partial_0 \oper(\sigma_1\Delta\apfi) =
2\left(\dweight\right) \gamma^0\Sigma_1 C'
\\
\gamma^2\partial_2 \oper(\sigma_1\Delta\apfi)=
2\left(\dweight\right) \gamma^2\Sigma_1 C'
\end{array}
\ed
There is no $x_3$ dependence on the right hand of (\ref{fftjes1}). 
Moreover, when $\gamma^0\partial_0 \oper(\sigma_1\Delta\apfi) = \gamma^2\partial_2 \oper(\sigma_1\Delta\apfi)$, the term  $2\left(\dweight\right)$ drops of from both sides.
This subsequently,  implies,
\bg\label{fftjes2}
\gamma^0\Sigma_1 C'
=
\gamma^2 \Sigma_1 C'
\ed
Note that the $m^5$ containing term drops off as it is clear from (\ref{fftjes2}).
It is found
This implies that on the left hand of (\ref{fftjes2}) 
\bg\label{fftjes5}
\gamma^0\Sigma_1
C'
=
\left(
\begin{array}{cc}
1 & 0
\\
0 & -1
\end{array}
\right)
\left(
\begin{array}{cc}
0 & -\sigma_3(1+\sigma_1)
\\
-\sigma_3(1+\sigma_1) & 0
\end{array}
\right)
C'
\ed
This can be evaluated as
\bg\label{eval1}
\gamma^0\Sigma_1
C'
=
\left(
\begin{array}{cc}
0 & -\sigma_3(1+\sigma_1)
\\
\sigma_3(1+\sigma_1) & 0
\end{array}
\right)
\left(
\begin{array}{c}
C'_\upsilon
\\
C'_\delta
\end{array}
\right)
=
\left(
\begin{array}{r}
-\sigma_3(1+\sigma_1)C'_\delta
\\
\sigma_3(1+\sigma_1)C'_\upsilon
\end{array}
\right)
\ed
On the right hand of (\ref{fftjes2})
\bg\label{fftjes6}
\gamma^2\Sigma_1
=
\left(
\begin{array}{cc}
0 & \sigma_2
\\
-\sigma_2 & 0
\end{array}
\right)
\left(
\begin{array}{cc}
0 & -\sigma_3 (1+\sigma_1)
\\
-\sigma_3 (1+\sigma_1) & 0
\end{array}
\right)
=
\\\nonumber
=
\left(
\begin{array}{cc}
-\sigma_2\sigma_3 (1+\sigma_1) & 0
\\
0 & \sigma_2\sigma_3 (1+\sigma_1)
\end{array}
\right)
\ed
Then, let us explicitly look at the $4\times 1$ vector $C'$.
We have the entries like  $C'^T = (\seupo,\seupt,\sedoo,\sedot)$.
And note that we, at this point, are still free to have $\seupo +\seupt \neq 0$ and $\sedoo + \sedot \neq 0$.
Therefore,
\bg\label{ok2}
\gamma^2 \sigma_3(1+\sigma_1)C' =
\left(
\begin{array}{r}
-\sigma_2\sigma_3(1+\sigma_1)\seeup
\\
\sigma_2\sigma_3(1+\sigma_1)\sedown
\end{array}
\right)
=
\sigma_3
\left(
\begin{array}{r}
\sigma_2(1+\sigma_1)\seeup
\\
-\sigma_2(1+\sigma_1)\sedown
\end{array}
\right)
\ed
Looking at the upper two of the $4\times 1 $ vector $C'$ and equate (\ref{eval1}) and (\ref{ok2}),
i.e. $-\sigma_3(1+\sigma_1)C'_\delta=-\sigma_2\sigma_3 (1+\sigma_1)C'_\upsilon$
we get:
\bg\label{up21}\nonumber
-(1+\sigma_1)
\left(
\begin{array}{c}
\sedoo
\\
\sedot
\end{array}
\right)
=
\sigma_2
(1+\sigma_1)
\left(
\begin{array}{c}
\seupo
\\
\seupt
\end{array}
\right)
\ed
Recall the $\sigma$ matrices in (\ref{gg1}) and acknowledge
\bg\label{up22}
\left(
\seupo +\seupt
\right)=
i\left(
\sedoo + \sedot
\right)
\\\nonumber
\left(
\seupo +\seupt
\right)=
-i\left(
\sedoo + \sedot
\right)
\ed
It gives $\seupo+\seupt = \sedoo +\sedot =0$.
\subsection{Consequence of $\seupo+\seupt = \sedoo +\sedot =0$.}\label{cons}
Let's inspect $\Gamma'C'$.
And this leads to $(\pm 1/2)(\sigma_1-\sigma_3)(1+\sigma_1)C'_{\upsilon(\delta)}$.
Hence, for $C'_\upsilon$
\bg\label{cons1}
\frac{\pm 1}{2}
\left(
\begin{array}{rr}
-1 & 1
\\
1 & 1
\end{array}
\right)
\left(
\begin{array}{cc}
1 & 1
\\
1 & 1
\end{array}
\right)
\left(
\begin{array}{r}
\seupo
\\
\seupt
\end{array}
\right)
\ed
That gives
\bg\label{next}
(\pm 1)
\left(
\begin{array}{cc}
0 & 0
\\
1 & 1
\end{array}
\right)
\left(
\begin{array}{c}
\seupo
\\
\seupt
\end{array}
\right)
=
(\pm 1)
\left(
\begin{array}{c}
0
\\
\seupo+\seupt
\end{array}
\right)
\ed
Because, $\seupo+\seupt = 0$, the $(\Gamma'C')_\upsilon = 0$.
Similarly, $(\Gamma'C')_\delta = 0$ for $\sedoo +\sedot =0$.
Hence, $\Gamma'C'=0$ which follows from adding $m\sigma_1\Delta\apfi$ to $\apfi$  and the approximation neglect $O(m^4)$ of equation (\ref{KG4}).
\section{Discussion}
In our paper we employed a second perspective that may "observe" the ultra light neutrino. 
It is essentially a weighted $x_2$ and $x_0$ exchange of role, but is different from (\ref{psihat}) in section-\ref{cconj}. 
Then, when a mixed coss-hair Klein-Gordon like equation  (\ref{eqbas}) is physically true, it forces the $\Delta\apfi = 0$.
In other words, it is impossible for a neutrino with $\Delta \apfi \neq 0$ to exist in the cross-hair equation (\ref{eqbas}).   
But then $\Delta\apfi \neq 0$ gives a valid solution to $\dirm$. 
It is noted that the mixed equation (\ref{eqbas}) is Lorentz invariant.

In the Dirac quantization in the neglect of term $O(m^4)$ case, there could be a physics reason why a neutrino with a $\Delta\apfi\neq 0$, may not be in
the cross-hairs of two perspectives with respective metric signatures $(+,-,-,-)$ and $(-,-,+,-)$. 
\subsubsection{How from $g$ to $\hat{g}$}
A next intriguing question could be what may possibly change a metric tensor from $g^{\mu,\nu}$ to $\hat{g}^{\mu,\nu}$ 
in an almost free neutron environment?
It was already noted in section-\ref{cconj}, that, mathematically, the charge conjugation matrix can involved in going from $\barpar$ to $\hatbp$ for a reduced zero mass Dirac equation.
However, both $x_0$ and $x_2$ are real in our use of $\hatbp$.

We try to answer the question via a perturbation of the metric tensor $g^{\mu.\nu}$ with entries from a sort of Schwarzschild metric.
The Schwarzschild metric tensor in cartesian coordinates can be approximated  viz.
\cite[equation (10.38)]{Schutz}.
Suppose now $M$ is a relatively large mass, then,
$g_S^{0,0}=(1-\frac{2M}{r})$ and $g_S^{k,k}=-(1+\frac{2M}{r})$, with, $r=\sqrt{\sum_k x_k^2}$ together with $k =1,2,3$ and $g_S^{\mu,\nu}=0,\, \mu \neq \nu$.
And so we may hypothesize that initially $\hat{g}_{\text{tr}}^{2,2}=g_S^{2,2}g^{2,2}=-(1+\frac{2M}{r})g^{2,2}$ while $\hat{g}_{\text{tr}}^{0,0}=-g_S^{0,0}g^{0,0}=-(1-\frac{2M}{r}) g^{0,0}$ and that subsequently we may see $\hat{g}_{\text{tr}}^{2,2} \rightarrow \hat{g}^{2,2}$ and $\hat{g}_{\text{tr}}^{0,0} \rightarrow \hat{g}^{0,0}$.
It is observed that when the operator 
\bg\nonumber
\hattr \, =\hat{\gamma}^2\left(1+\frac{2M}{r}\right)\partial_2 -\hat{\gamma}^0\left(1-\frac{2M}{r}\right)\partial_0 -\gamma^1\partial_1 -\gamma^3\partial_3
\ed
instead of $\hatbp$ is employed in (\ref{eqbas}), the equation $\gamma^0\partial_0 \phi_{1,3} = \gamma^2\partial_2 \phi_{1,3}$ in (\ref{KG4}) remains valid.
The $-i\frac{2M}{r}(\gamma^2+\partial_2 +\gamma^0\partial_0)\barpar \varphi$, with $M\gg m$, from the term $-\hattr\,\,\,\, \barpar \varphi$, cancels with the $(-im)\frac{2iM}{r}(\gamma^2\partial_2 +\gamma^0\partial_0) \varphi$ from the $-im(\hattr\,\, - \barpar)\varphi$ term.

In the construction of the cross-hair equation (\ref{eqbas}), it appears possible to go to $\hatbp$, via $\hattr$\, , starting from $\barpar$.
The product transformation such as e.g. $-(1+\frac{2M}{r})g^{2,2}$ is hypothetical.
It could be a first approximation of a physics underpinning of the possibility of two perspectives that simultaneously can have a neutrino in the cross-hairs (\ref{eqbas}). 
If $\hatbp$ and $\barpar$ do not physically go together then the Lorentz invariance of (\ref{eqbas}) may suggest otherwise. 
\subsubsection{Extension of $\hattr$}
If we add to $\hat{g}_{\text{tr}}$ additional non-diagonal terms  $\delta\hat{g}_{\text{tr}}$,  like e.g. $\hat{g}_{\text{tr}}^{0,2}=\frac{2M}{r}$ and $\hat{g}_{\text{tr}}^{2,0}= -\frac{2M}{r}$, the determinant $\det(\hat{g}_{\text{tr}})$ will remain, $-1$. In this off-diagonal perspective case, the equation $\gamma^0\partial_0 \phi_{1,3} = \gamma^2\partial_2 \phi_{1,3}$ of (\ref{KG4}) will remain valid as well.
It is likely that in this case the perspective with $\hat{g}_{\text{tr}}$ is something physical.

The following may perhaps be an interesting aside. 
If only $x_0$ and $x_2$ variables and zero mass is actually what we are looking at, the charge conjugation matrix $\mathcal{C}$ plus the exchange of $x_0$ to a space and $x_2$ to a time will have the result of going from $g$ to the $\hat{g}$ perspective.
\section{Conclusion in a question format}
How is it possible in empirical reality that a $\hhoed$ and $\barpar$ mixed equation, i.e. $\hhoed$ replaces $\hatbp$ in (\ref{eqbas}), is not Lorentz invariant but refers to the same Klein-Gordon type of equation, \emph{while}, a mixed $\hatbp$ and $\barpar$ equation is Lorentz invariant but 
the $\hatbp$ does not give the same Klein-Gordon like equation as the $\barpar$. 
Moreover if we have, $ \Delta \apfi \neq 0$, 
the cross-hair equation (\ref{eqbas}) is contradictory to an $O(m^4)$ approximation of a neutrino Dirac equation.
This is so despite $(i\hatbp + m)(i\barpar -m) \varphi (x) = 0$ is Lorentz invariant. 
The \emph{question} then is:  
Do (approximately free) neutrinos align themselves to the coexistence of a second perspective by adjusting to a state with $\Delta\apfi = 0$.
That is, a state where only separable variables exist.

Of course, when there is e.g. no $m\sim 10^{-46}$ [kg] $\sim  
10^{-11} eV$, light neutrino, or the two perspectives do not go together, this is a kind of theoretical Ehrenhaftian  "Dreckeffect" \cite[page 39]{Fey}. 
But note, the $10^{-11} eV$ neutrino appears cosmologically allowed \cite[page 139]{Kolb}.
\bibliography{apstemplate}

@PREAMBLE{
 "\providecommand{\noopsort}[1]{}" 
 # "\providecommand{\singleletter}[1]{#1}%" 
}

@BOOK{Bjork,
   author       = {J. D. Bjorken and S. D. Drell},
   year         = 1964,
   title        = {Relativistic Quantum Mechanics},
   publisher    = {McGraw Hill}
}

@BOOK{Fey,
   author       = {P. Feyerabend},
   year         = 1978,
   title        = {Against Method},
   publisher    = {Verso, London, UK}
}

@BOOK{Schutz,
   author       = {B. Schutz},
   year         = 2009,
   title        = {A first course in general relativity},
   publisher    = {Cambridge University Press}
}

@BOOK{Kolb,
   author       = {E.W. Kolb and M.S. Turner},
   year         = 1989,
   title        = {The Early Universe},
   publisher    = {Addison Wesley Publishing}
}

@ARTICLE{Ander,
author = {C. D. Anderson},
year = 1933,
journal = {Phys. Rev.},
 volume= "43",
 pages="491",
}

@ARTICLE{Bryan,
   author = "B. Sanctuary",
   year   = "2024",
   journal= "Mathematics",
   volume = "12",
   pages  = "1962",
}

@ARTICLE{Sach,
   author       = "A. Sakharov",
   year         = "1984",
   journal      = "JETP ",
   volume       = "87",
   pages        = "375",

}
\end{document}